# Offloading Content with Self-organizing Mobile Fogs

Junaid Ahmed Khan,*‡ Cedric Westphal* and Yacine Ghamri-Doudane†

*University of California, Santa Cruz, CA, USA
& Huawei Technology, Santa Clara, CA, USA

‡Univ Lyon, INSA Lyon, INRIA, CITI Lab, Villeurbanne, France

†L3i Lab, University of La Rochelle, France

junaid.khan@inria.fr, cedric@soe.ucsc.edu, yacine-ghamri@univ-lr.fr

*Abstract*—Mobile users in an urban environment access content on the internet from different locations. It is challenging for the current service providers to cope with the increasing content demand from a large number of collocated mobile users. In-network caching to offload content at nodes closer to users alleviate the issue, though efficient cache management is required to find out who should cache what, when and where in an urban environment, given nodes limited computing, communication and caching resources. To address this, we first define a novel relation between content popularity and availability in the network and investigate a node's eligibility to cache content based on its urban reachability. We then allow nodes to self-organize into mobile fogs to increase the distributed cache and maximize content availability in a cost-effective manner. However, to cater rational nodes, we propose a coalition game for the nodes to offer a maximum "virtual cache" assuming a monetary reward is paid to them by the service/content provider. Nodes are allowed to merge into different spatio-temporal coalitions in order to increase the distributed cache size at the network edge. Results obtained through simulations using realistic urban mobility trace validate the performance of our caching system showing a ratio of $60-85\%$ of cache hits compared to the $30-40\%$ obtained by the existing schemes and $10\%$ in case of no coalition.

## I. Introduction

The increase in the number of smart mobile devices resulted in the growth of content demand by lots of consumers in closer urban proximity, some even using multiple portable devices. For example, a large number of mobile users in an urban environment can be interested to watch the video of a latest episode of a popular TV show/drama or sports highlights. Provisioning of such popular content to each user requires lots of redundant connections between users and the service provider, given that the content is requested by lots of spatio-temporally co-located users with similar social interests.

Nowadays, it becomes more and more challenging for the current "connection-centric" network infrastructure to facilitate content availability for such large number of mobile users in close proximity while offering attractive tariff plans with no bandwidth limitations. One of the solutions advocated by the research community is the use of Content-Centric Networking (CCN) [1] paradigm which addresses the issue by decoupling the content provider and consumer along in-network caching at intermediate nodes. Content caching at individual nodes exploit different content replacement strategies such as First-In First-Out (FIFO), Least Recently Used (LRU) and Least Frequently Used (LFU). However, there is a need to go beyond individual caching policies towards a collaborative content caching approach near mobile users.

In this paper, we target the content caching at the edge problem where nodes in an urban environment can subscribe to offer distributed caches to a service provider [2]. We define *Mobile fogs* as set of co-located mobiles nodes that can self-organize to offer distributed resources (computing, communications, caching) as an ad-hoc network at the network edge. The aim is to allow nodes to collaborate in order to increase the "virtual cache" closer to mobile users, maximize content availability and minimize cost for the service provider as well as the users. This however invokes the following questions; first, given the massive content constantly generated and consumed by mobile devices [3], which content is important to cache with respect to the social interests of geographically co-located users at different locations and times of the day? Second, which nodes can be considered suitable candidates for caching, given limited bandwidth and monetary resources, both, for the service provider and the mobile nodes? Third, given the existence of lots of such subscribed nodes including rational nodes, how to incentivize mobile users and compensate for their incurred cost of caching content at different urban neighborhoods, specially, in populated areas (spatial) or during peak traffic hours (temporal) to maximize content availability?

Thus, it is challenging to address questions regarding "Who should cache what, when and where?", requiring an understanding of the content as well as the node's spatio-temporal profiles. To address this, inspiring from social networks metrics (centrality), we first define a novel relation considering simultaneously the content spatio-temporal popularity and availability to characterize its importance. Based on the content profile, a node autonomously finds its caching capability by computing its local [4] centrality as a measure of its connectivity and reachability in the network. The existence of rational caching nodes cannot be ignored, where such users opt to reduce the consumption of their resources such as battery and storage dedicated for caching. Thus, we present a coalition game to incentivize co-located nodes already interested in the content to merge into coalitions and collaboratively offload a maximum amount of content locally to other nearby users. Nodes in coalition receive a monetary reward proportional to the amount of content offloaded in a neighborhood where coalitions or mobile fogs formed by all subscribed nodes in each neighborhood can be considered as the grand coalition providing a larger city-wide distributed virtual cache. Since the distributed caching problem is NP-Hard [5], an optimization problem is thus formulated with a heuristic to maximize the offered distributed cache in the network. The contribution of this paper can be summarized as follows:

- A novel content spatio-temporal availability-popularity relation is proposed to decide a content importance in order to be cached by nodes at the network edge.
- A new node eligibility metric is presented for social-aware content caching. It classifies a node's caching capability proportional to its relation to content, connectivity, cost and reachability in the network.
- A coalition game is proposed to incentivize caching nodes, allowing a service provider to optimally select nodes as caches in an urban environment, while compensating for their costs and maximizing content availability in different locations and times.
- A distributed algorithm is proposed for nodes with caching capability to self-organize and collaboratively provide a maximum virtual cache near end users at different locations and times.

In order to validate our approach, we compare the cache hits and offload benefit of mobile fogs formed with state of the art node centrality schemes as well as a case when no coalition is formed. Results show that our approach outperforms such approaches as well as the case without coalitions by substantially offloading the content at the distributed mobile fogs while achieving around $60\%$ of cache hits compared to the $30-40\%$ obtained by the existing schemes and $10\%$ in case of no coalition.

The remaining of the paper is organized as follows. Section II discusses the related work. In Section III we define the metrics for the content and node eligibility. The coalition game for the formation of mobile fogs along the optimal coalitions selection is described in Section IV. In Section V, we discuss the performance evaluation and results. Section VI concludes the paper along insights into future directions.

## II. RELATED WORK

Content Caching is studied for some time by the research community mainly in Small Cell Networks (SCNs) [6], [7], Content Distribution Networks (CDNs) [8] and Information/Content Centric Networking (ICNs/CCNs). For example, in [9], distributed cache management decisions are made in order to efficiently place replicas of information in dedicated storage devices attached to nodes of the network using ICN. Similarly [10] address the distribution of the cache capacity across routers under a constrained total storage budget for the network. A sub-optimal heuristic method is proposed based on node centrality for the case of frequent content publishing. The authors found that network topology and content popularity are two important factors that affect where exactly should cache capacity be placed. In a recent work [11], game theory is exploited for caching popular videos at small cell base stations (SBSs). A Stackelberg game model is presented for the service providers to lease its SBSs to video retailers in order to gain profits as well as reduce the costs for back-haul channel transmissions. Another work, [12] proposed a game theoretic approach in ICN to stimulate wireless access point owners to jointly lease their unused bandwidth and storage space to a content provider under partial coverage constraints.

An example of social-aware caching approach in Delay Tolerant Networks (DTNs) is presented in [13] where the authors proposed a cooperative caching scheme based on the social relationship among nodes in DTNs. Content is dynamically cached at selective locations (i.e. nodes along the common request forwarding paths) in the network with the highest social levels identified as cluster head nodes. A cache replacement policy based on the content popularity is presented as a function of both the frequency and recency of data access. Similarly, Socially-Aware Caching Strategy (SACS) [14] for Content Centric Networks uses social information in order to privilege influential users in the network by pro-actively caching the content they produce. The authors detect the influence of users within a social network by using the Eigenvector and PageRank centrality measures. The above mentioned caching schemes rely on typical centrality measures which are unsuitable and unstable in dynamic connectivity scenarios in an urban environment. Moreover, existing schemes considers only content popularity along policies such as FIFO, LRU an LFU, which do not completely capture the spatio-temporal content characteristics to address the requirements of urban mobile users. Thus, none of existing approach is able to model the content profile with respect to the economic and social interest of large (scalable) number of users.

We address the caching problem to find out who should keep what and how, jointly with the existence of rational caching nodes. Unlike the above mentioned approaches, (i) we derive a relation between spatio-temporal content availability and popularity allowing a node to decide its caching decisions. It considers the extent it is related to the content based on the frequency of interest it satisfied for the content. (ii) We propose novel content caching metrics which go beyond typical FIFO, LRU, LFU schemes towards a collaborative strategy, while simultaneously considering recency, frequency as well as accounting for large number of users with socially similar interests. (iii) Our popularity metrics are adapted to urban mobile environments considering node local connectivity (topology) as well as its reachability and (iv) the mobile fog formation ensures maximum content availability with the best candidate nodes in coalition at different locations and times, while, at the same time incentivizing nodes. To the best of our knowledge, there exist no scheme to provide such a scalable content caching model to offload traffic at user devices including mobile nodes in urban environment.

## III. WHO SHOULD CACHE WHAT?

The first step towards efficient content caching is to understand the eligibility of the content to be cached based on a relevant amount of user demands relative to the content as well as its availability in the network. Before discussing the content importance, we first present the system model to consider for the rest of the paper.

### A. System Model

*1) Business Model:* We consider a service provider and a set of subscribed users requesting content at different locations and times. There are three players involved with their respective business gains and costs, (i) the service provider whose benefit is proportional to the amount of content downloaded by the subscribed users. (ii) User nodes with sufficient resources (communication, caching, computing, energy etc.) interested in the content downloads it and are capable to cache it at their fixed access points, small-cells or even mobile devices in order to provide it to other nearby users. However, caching and delivering content incur cost for such users proportional to their provided resources such as battery and storage. Therefore, they are paid a monetary reward by the service provider proportional to the amount of content delivered to other users where such reward is more than their consumed resources and less than the cost of using the infrastructure network. (iii) The user node subscribed to consume the content with no caching capability, thus paying an amount to the service provider proportional to the downloaded content. The cost for a user to download content from a nearby user in an ad-hoc network is less than using the infrastructure network.

Caching content at the edge and allowing users to use the low cost ad-hoc network to retrieve content locally from nearby nodes is economically beneficial for all the three players; (a) The reduction in service provider cost and bandwidth for not using the infrastructure network. (b) The reward paid to the nodes caching content, greater than their incurred cost, and (c) the reduction of cost for the user downloading content from nearby nodes in an ad-hoc network.

*2) Network Model:* We consider an ad-hoc network modeled as an undirected graph $G(\mathbb{V}(t), \mathbb{E}^v(t))$, where $\mathbb{V} = \{v\}$ is the set of nodes and where the time instant $t$ integrates the temporal network characteristics. $\mathbb{E}^v(t) = \{e_{jk}(t) \mid v_j, v_k \in \mathbb{V}, j \neq k\}$ is the set of edges $e_{jk}(t)$ modeling the existence of a communication link between nodes $j$ and $k$ at time instant $t$. To consider the dynamic nature of the network topology, we assume the time $T = (\overline{t_1}, \overline{t_2}, ...)$ as a sequence of regular time-slots, where the $k^{th}$ time-slot is represented as $\overline{t_k} = [t_k, t_{k+1})$. Similarly, the urban environment is represented by the undirected graph $G(\mathbb{L}, \mathbb{E}^l)$, the set of vertices $\mathbb{L} = \{l\}$ represents different locations $l$ and the set of edges $\mathbb{E}^l = \{e_{pq} \mid l_p, l_q \in \mathbb{L}, p \neq q\}$ are the respective boundaries that connects different neighborhoods. It is to note that there is a precision versus overhead trade-off, i.e. The more locations and time-slots, the more computing overhead. The granularity of such spatio-temporal division depends on the application requirements. To provide a generic solution, we consider a generalized location $l \in L$ and time-slot $\overline{t} \in T$ in the remaining of the paper.

Table I: List of Notations

| Notation | Description |
|---|---|
| $\mathbb{V}$ | Set of nodes |
| $\mathbb{L}$ | Set of locations |
| $\mathbb{X}$ | Set of content |
| $\mathbb{E}^l / \mathbb{E}^v(t)$ | Edges set between locations/nodes at time $t$ |
| $\mathbb{E}$ | Edges set between nodes and locations |
| $T/t/\overline{t}$ | Total time / time instant / Time slot |
| $t_v^{x_f}$ | Time instant: last interest satisfied |
| $\overline{t_v^{x_{av}}}$ | Time-slot: average content validity |
| $\overline{t_k}$ | $k^{th}$ time-slot under consideration |
| $\lambda$ | Number of received interests for content $x$ |
| $\Lambda$ | Number of received interests at $v$ for all contents |
| $M$ | Total number of nodes |
| $N$ | Total number of content chunks |
| $A_v^x$ | Content $x$ availability at node $v$ |
| $Q_x$ | Number of chunks for content $x$ |
| $F_x$ | File size for content $x$ |
| $F_v$ | File size for all content at $v$ |
| $p_x$ | User interest probability for content $x$ |
| $\tau_v^x$ | Content validity metric |
| $\delta$ | Content replacement tuning parameter |
| $I_v^x$ | Interest satisfaction frequency |
| $r_v/R_v$ | Successful/Overall interests satisfied |
| $\psi_v^x$ | Content popularity/availability relation |
| $C_v$ | Node centrality |
| $f_I^v$ | Content importance function |
| $f_{L,T}^v$ | Spatio-temporal availability function |
| $MI_v(l, \overline{t_k})$ | Mutual information between location & time |
| $P_v(l, \overline{t_k})$ | Joint probability between location & time for node |
| $P_v(l), P_v(\overline{t_k})$ | Joint probability of location & time for node |
| $f_\Gamma^v$ | Neighborhood connectivity |
| $B_v/B_v^t$ | Available/total buffer at node $v$ |
| $f_v^C / f_v^R$ | Node cost/reward function |
| $C_d$ | Costs incurred by node |
| $R_d$ | Reward offered to node |
| $U_v/U_s$ | Node/coalition utility function |
| $S$ | Set of coalitions |
| $\succ$ | preference relations among coalitions |
| $y_v$ | node payoff in imputation |
| $w_s$ | weight for coalition $s$ |

We define the set of content as $\mathbb{X} = \{x\}$ where $x$ is an indivisible content chunk in the network. We need to take into account a spatio-temporal relation of its availability versus popularity at a location $l \in \mathbb{L}$ and a time slot $\overline{t}$. Therefore, in the remaining of this section we model and map the content availability and its popularity with respect to location and time.

### B. Spatio-temporal Content Profile

It is important to consider both the spatial and temporal characteristics of the content available as well as demanded in the network. Therefore, we let the node extrapolate from the past information for the cached content availability and popularity to decide regarding its future profile. We define below the proposed content availability and popularity respectively:

***Definition 1***: (Content Availability) We formally define the spatio-temporal availability for a content piece $x \in \mathbb{X}$ cached at a node $v \in \mathbb{V}$ as $A_v^x(l, \bar{t}) = \begin{cases} 1, & \text{x cached at v} \\ 0, & \text{otherwise} \end{cases}$, where $\bar{t}$ is the time slot at which the content is available at the node $v$ at a location $l \in \mathbb{L}$. Content availability differs with respect to time and location, similarly, the content availability is zero for all the content $x \in \mathbb{X}$ not cached at the node at a particular location and time slot.

To model the storage requirements for a particular content $x$, we denote $Q_x$ as the number of chunks/pieces cached at a node, assuming the possibility of having multiple content chunks. The file size for the content can be defined as $F_x = Q_x \cdot |x|$. Thus, the file size of all the content cached at a node $v$ is given as $F_v(l, \bar{t}) = \sum_x A_v^x(l, \bar{t}) \cdot F_x$, i.e. the sum of file sizes for all the possible content at location $l$ and time slot $\bar{t}$.

The content popularity can be shared with the nodes using three approaches, (i) offline method by the content operator as a control message; (ii) part of content header shared by the service provider; and (iii) local monitoring by the nodes taking into account the number of user interests for the content. (i) and (ii) requires global knowledge only available to the service provider. We define below (iii) for a node, the cached content spatio-temporal popularity with respect to a location $l$ and time-slot $\bar{t}$ by local monitoring.

***Definition 2***: (Content Popularity) The spatio-temporal content popularity at a node is represented by the probability $p_x(l, \bar{t}) = \frac{\lambda(l, \bar{t})}{\Lambda}$ as a measure of the user interests for content $x$ at location $l$ and time $\bar{t}$.

Here, $\lambda(l, \bar{t})$ represents the number of interest for the content $x$ at location $l$ and time $\bar{t}$ and $\Lambda = \sum_{X_v} \lambda(l, \bar{t})$ is the total interests for all contents cached $X_v$ at the node $v$ at the required location $l$ and time slot $\bar{t}$. The above definition considers varying content popularity at different times and locations, thus modeling the existence of users with different social interests in different neighborhoods.

It is important to find the node relation to the cached content with respect to the received user interests in order to efficiently cache content of interest for the intended users in the network. Therefore, we define below a user interest satisfaction based metric which complies with such criteria:

***Definition 3***: (Interest Satisfaction Frequency) We define $I_v^x(l, \bar{t}) = \frac{r_v(l, \bar{t})}{R_v}$ as the frequency of user interests satisfied by the node for content $x$ at location $l$ and time-slot $\bar{t}$, where $r_v(l, \bar{t})$ are the number of successful responses in the previous time slot and $R_v = \sum_{X_v} r_v(l, \bar{t})$ are the overall successful responses for all the contents $X_v$ cached at the node $v \in \mathbb{V}$ at the respective location and time slot.

We assume each user interest specifies a hint (such as Time To Live - TTL) as the content validity for the intended user. Since such TTLs vary for interests received from different users at the node $v$. We let the node compute the average interest validity hint from the received user interests. It is defined as the average time slot specified by the user interest as the cached content validity deadline, averaged over all the interests received in the past for the content. It enables the node to decide whether the cached content is still demanded by users in the network or not. Therefore, we allow each node to devise a content replacement strategy based on such user specified content validity hint.

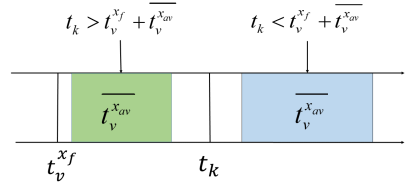

Figure 1: Content validity hint indicated by user interests

Let $t_v^{x_f}$ be the time instant when the interest for the content $x$ was previously satisfied by node $v$ and the average interest validity time-slot is represented as $\overline{t_v^{x_{av}}}$ for the content.

***Definition 4***: (Content Validity) Leveraging the user interest validity hint, we define $\tau_v^x(l, \bar{t})$ as the measure of the content validity scope for the next time instant $t_{k+1}$. Different functions can be used, however we use here the commonly used exponential decay function:

$$\tau_v^x(l, \bar{t}) = \begin{cases} 1 & t_k \leq t_v^{x_f} + \overline{t_v^{x_{av}}} \\ e^{-\delta(t_k - \overline{t_v^{x_{av}}})} & t_k > t_v^{x_f} + \overline{t_v^{x_{av}}} \end{cases}$$

where, $\delta$ is the tuning parameter to adjust the exponential decay depending on the application requirements as in Figure 1. For the current time instant $t_k$ less/greater than the sum of the time instant the content was last responded $t_v^{x_f}$ and the time slot $\overline{t_v^{x_{av}}}$ specifying the average user interest validity.

In case the node receives no active interest in the previous time slot and the average interest validity is expired, the content validity follows an exponential decay since the information is of less importance in the network. On the other hand, $\tau$ is set to unity for the content required to be constantly cached at the node.

***Definition 5***: (Content Availability vs Popularity Relation) We present a metric on the content importance for a node to model together a cached content's "availability" while considering its "popularity" with respect to the received user interests. The content availability at a location $l$ and time $\bar{t}$ can be mapped together with its popularity in the network by the relation:

$$\psi_v^x = p_x \cdot \tau_v^x \cdot I_v^x, \qquad A_v^x \neq 0 \qquad (1)$$

where, $A_v^x$ is a binary variable modeling the content $x$ availability at a node $v$. For a cached content i.e. $A_v^x \neq 0$, Equation 1 allows nodes to estimate the cost and reward for caching and delivering content as well as allowing nodes to autonomously decide content replacement decisions. For a given content set $X_v$, it can classify content by looking into the user interests ($p_x$) and find out how long it needs to keep the content

($\tau_v^x$). Similarly, the node can find the degree of relevance with respect to cached content from the amount of interest it already satisfied for the content ($I_v^x$). At the same time, $\psi$ ensures diversity in caching content as there might be content not popular overall, though more related to the individual node (high $I_v^x$), and therefore, should be kept at the node. It is to note that $\psi_v^x$ is a single metric that incorporates a hybrid of content recency, popularity and availability as a linear content eligibility metric at a node depending only on its local knowledge. At the same time, each term in Equation 1 consider content as a stable metric, independent of unstable metrics such as nodes interactions and mobility.

### C. Node Eligibility

We believe content caching should be made at social-aware mobile nodes identified in the network using centrality measures. The idea is, more precisely, to classify a node for content caching based on the social needs of users in an urban environment. However, it is not possible to compute typical centrality measures (Degree, Closenesss, Betweenness, Eigenvector) in a dynamic network topology for mobile nodes. Furthermore, existing schemes follow a network-centric approach requiring complete network connectivity in order to analyze the network, thus, ignoring the content importance with respect to the user relevance. Similarly, using the nodes contact frequency and duration to decide its capability to cache in the network is a challenging task because of the rapid changes in the time evolving network topology due to mobility. To overcome this, we propose the use of a novel social aware metric, which simultaneously considers three essential parameters, the content user relevant importance, the node spatio-temporal availability and its network connectivity [4] allowing it to find its local caching capability without affected by its mobility.

- Content importance: It measures the cached content eligibility $\psi$ (Equation 1).
- Spatio-temporal availability: reflects the node's recursive presence in a particular location $l$ at a time-slot $\bar{t}$, reflecting its correlation to the location. It is the mutual information the location $l$ and time-slot $\bar{t}$ share about the node expressed as:

$$MI_v(l; \bar{t}) = \sum_{\forall l \in \mathbb{L}} \sum_{\forall \bar{t} \in T} P_v(l, \bar{t}) \log \left( \frac{P_v(l, \bar{t})}{P_v(l) P_v(\bar{t})} \right), \quad (2)$$

where, $P_v(l, \bar{t})$ is the joint and $P_v(l)$ and $P_v(\bar{t})$ are the marginal probabilities of the node presence at the location $l$ and time-slot $\bar{t}$ respectively, usually obtained prior to the eligibility computation time using different localization measures.
- Neighborhood importance: Neighborhood importance considers the node topological connectivity. Different metrics can be used to to find a node's local connectivity, we use an easier to compute average degree (number of neighbors) $k_v$ for a node $v$ observed in a prior time-slot.

The nodes first classify the content using the relation $\psi_v^x$ in Equation 1 and then use the content eligibility to find its relative importance in the network as a local information hub responsible for the efficient caching. The local centrality is represented as $C_v$, computed at a time instant $t_k$:

$$C_v(t_k) = f_I^v(t_k) \cdot f_{L,T}^v(t_k) \cdot f_\Gamma^v(t_k) \quad (3)$$

where, $f_I^v$, $f_{L,T}^v$ and $f_\Gamma^v$ are the content importance, node's spatio-temporal availability and its neighborhood, respectively. Thus, a node more related to a content (high $f_I^v$), more frequently present at the location $l$ and time-slot $\bar{t}$ (high $f_{L,T}^v$) and better connected (high $f_\Gamma^v$) is the best candidate to cache the content in that location and time-slot.

*Node Utility:* The node utility is composed of its reward and associated cost proportional to the amount of content delivered to users. The node cost function is defined as:

$$f_v^C(l, \bar{t}) = \frac{1}{|\mathbb{X}|} \sum_x d_v^x(l, \bar{t}) \times C_d(l, \bar{t}), A_v^x \neq 0, \forall x$$

where, $d_v^x$ is the amount of content chunks delivered and $C_d$ is the cost per content chunk delivery incurred to the node. The monetary reward paid to the node for caching besides the natural incentive for the node to be itself interested in the content is given by the function $f_v^R(l, \bar{t}) = d_v^x(l, \bar{t}) \times R_d$. It is proportional to the content delivered, where $R_d$ is the reward paid per delivered content chunk dedicated by the service provider. We derive the node's utility as:

$$U_v(l, \bar{t}) = f_v^R(l, \bar{t}) - f_v^C(l, \bar{t}),$$

where, $f_v^R(l, \bar{t})$ and $f_v^C(l, \bar{t})$ are the node's reward and cost functions, respectively. Thus, the node computes its utility based on the locally known information regarding the amount of delivered content, its cost and the reward per unit delivery.

## IV. Distributed Fogs Formation as Coalition Game

The nodes collaborate to offer their buffer to the service provider by forming mobile fogs for distributed caching at set of co-located nodes aiming to maximize their profit/payoff. To manage this, we use game-theoretic concept of coalition formation where the service provider optimally selects the best coalitions formed by nodes in all areas at all time slots. The coalition game $G(\mathbb{V}, U)$ for the set of nodes $\mathbb{V}$ targets to form the set of coalitions $S = \{s\}$, $s \subseteq \mathbb{V}$ as the respective mobile fogs. The real-valued function $U : 2^{|\mathbb{V}|} \to \mathbb{R}$ associates to each coalition $s$ a value $U_s(l, t)$ as the total payoff available to players in coalition $s$ at the location $l$ and time slot $\bar{t}$ in an urban environment.

We define below the preference relation for the nodes preferences to merge into a particular coalition yielding a larger amount of utility. The service provider uses the same preference relation to select the best coalition of nodes with the maximum utility among the set of coalitions formed in the network at different locations and time slots.

*Definition 6:* (Coalitions Preference Relation) For any node $v$, a preference relation $\succ$ is defined as a complete, reflexive and transitive binary relation over the set of all coalitions that node $v$ can possibly form $\{s \subseteq \mathbb{V} : v \in s\}$.

Each node prefers to merge to a coalition with larger utility to increase its reward. For example, given two coalitions $s_1$ and $s_2$ for the node $v$ to join, $s_1, s_2 \subseteq \mathbb{V}$, such that $v \in s_1$ and $v \in s_2$, $s_1 \succ s_2$ indicates that node $v$ prefers to merge with $s_1$ over $s_2$. The node adds its utility to the coalition it joins, thus increasing the coalition utility.

The optimization problem can thus be formulated for the service provider as follows to choose the coalition (form mobile fogs) offering the maximum utility over preferences at different locations and times.

$$\begin{aligned} \underset{S}{\text{maximize}} \quad & \sum_{s \in S} U_s(l, \bar{t}) \qquad \forall v, \forall l, \forall \bar{t} \\ \text{subject to} \quad & \sum_{v \in s} A_v^x(l, \bar{t}) \geq A_v^{x_{\text{th}}}(l, \bar{t}), \forall s, \quad (C_1) \\ & \sum_{v \in \mathbb{V}} f_v^C(l, \bar{t}) \leq f_v^{C_{th}}(l, \bar{t}), \quad (C_2) \\ & f_v^R(l, \bar{t}) \geq f_v^{R_{\text{th}}}(l, \bar{t}), \quad (C_3) \\ & \max_{v \in \mathbb{V}} B_v(l, \bar{t}) \leq B_v^{\text{t}}(l, \bar{t}), \quad (C_4) \end{aligned}$$

The objective function maximizes the utility of coalitions formed by spatio-temporally co-located nodes offloading content for the service provider. The constraint $C_1$ states that the content availability in each coalition at each location and time should be more than a desired threshold specified by the service provider. It allow fairness by specifying a minimum content availability requirements at all nodes, locations and times. The constraint $C_2$ addresses the cost limitations where the cost paid to nodes as incentives is limited by a threshold cost a service provider can afford. Constraints $C_3$ and $C_4$ deal with the node individual resources, where, for all nodes in different coalitions, (i) the reward offered to the node for providing its storage should be greater than the associated cost and (ii) the buffer offered by the node should not exceed the total buffer size available at the node.

### A. Solution: The Core

The target is to motivate each node with cache to participate towards forming the grand coalition of all nodes also known as the core of the game i.e. the node receives

$$U_v(l, \bar{t}) = \begin{cases} U_v(l, \bar{t}), & \text{if joins coalition in l \& } \bar{t} \\ 0, & \text{otherwise} \end{cases}$$

To do so, we first define for each location and time slot an imputation $y = (y_1, y_2, ..., y_M)$, $M = |\mathbb{V}|$ as the distribution of the total payoff/reward among nodes such that (i) $y_v \geq U(\{v\}), \forall v$, i.e. a node receives at least as much as it could obtain on its own, without cooperating with anyone else (individual rationality). (ii) $\sum_v y_v = U(M)$ (efficiency). The core is the set of imputations $y$ such that $y_s = \sum_{v \in s} y_v \geq U_s, \forall s \subseteq \mathbb{V}$ and equality if $s = \mathbb{V}$.

*Non-emptiness and Uniqueness of the Core:* In order to derive the non-emptiness and uniqueness of the core, we define a set of weights $w_s$ for coalitions, where $0 \leq w_s \leq 1, \forall s \subseteq \mathbb{V}$, is a balancing set of weights if $\forall v \in \mathbb{V}, \sum_{s, v \in s} w_s = 1$.

**Algorithm 1** Self-organizing Coalition Formation
1: Initialize $s = \phi$,
2: **for** each location $l \in \mathbb{L}$, time-slot $\bar{t} \in T$ **do**
3: $\quad v' = \arg\max_{v \in \mathbb{V}} C_v(t_k)$
4: $\quad s = s \cup v'$
5: $\quad U_s \leftarrow U_s + U_{v'}$
6: $\quad$ **for** each content $x$ in $v'$ **do**
7: $\quad\quad A_s^x \leftarrow A_s^x + A_{v'}^{x_{\text{th}}}$
8: $\quad$ **end for**
9: $\quad f_s^C \leftarrow f_s^C + f_{v'}^C$
10: $\quad$ **while** $(f_s^C \leq f_s^{C_{th}})$ and $(A_v^x \geq A_v^{x_{\text{th}}})$, $\forall x$ **do**
11: $\quad\quad$ **for** each node $u'$, $u' \neq v'$ **do**
12: $\quad\quad\quad$ **if** $U_s > U_{u'}$ and $B_{u'} \leq B_{u'}^{\text{t}}$ **then**
13: $\quad\quad\quad\quad s = s \cup u'$
14: $\quad\quad\quad\quad U_s = U_s + U_{u'}$
15: $\quad\quad\quad\quad$ **for** each content $x$ in $u'$ **do**
16: $\quad\quad\quad\quad\quad A_s^x \leftarrow A_s^x + A_{u'}^{x_{\text{th}}}$
17: $\quad\quad\quad\quad$ **end for**
18: $\quad\quad\quad\quad f_s^C \leftarrow f_s^C + f_{u'}^C$
19: $\quad\quad\quad$ **end if**
20: $\quad\quad$ **end for**
21: $\quad$ **end while**
22: **end for**
23: **return** $s \in S, A_s^x, f_s^C, U_s$

*Definition 7:* (Balanced Coalition Game) The coalition game $(\mathbb{V}, U)$ is balanced if and only if, for every balancing set of weights $w$, we have $\sum_{\emptyset \neq s \subseteq \mathbb{V}} w_s \cdot U_s \leq U_{\mathbb{V}}$. We consider the weights for each coalition as the percentage of total nodes in the coalition. Then, we check whether the game is balanced by finding the condition in Definition 7. Thus, in a balanced game, any arrangement feasible for the different coalitions is feasible for the grand coalition as well. We can now state the Bondareva-Shapley Theorem as:

*Theorem 1:* (Bondareva-Shapley Theorem) The coalitional game $(\mathbb{V}, U)$ has a non-empty core if and only if it is balanced.

*Proof:* There can be at least one caching node in a coalition, thus i.e. $min(|s|) = 1, \forall l, \forall \bar{t}$. For each location and time, $^{\mathbb{V}}C_s$ combinations of nodes are possible in a coalition. The combinations of coalitions for all locations at all times can be given as $\prod_l \prod_{\bar{t}} {^{\mathbb{V}}C_s}(l, \bar{t})$. A single coalition in each location and time slot i.e. $^{\mathbb{V}}C_s = 1, \forall l, \forall \bar{t}$, results in $\prod_l \prod_{\bar{t}} {^{\mathbb{V}}C_s}(l, \bar{t}) = 1$. Thus, we are able to reduce $2^M$ possible combinations for coalitions to a single coalition, i.e. $1 << 2^M$, as the "grand" coalition.

### B. Algorithm: Optimal Selection among Coalitions

Algorithm 1 summarizes the coalition formation for distributed caching at spatio-temporally co-located nodes. For a location $l$ and time-slot $\bar{t}$, the service provider selects the node

$v' \in \mathbb{V}$ with the highest centrality as the delegate to locally cache content. Such high centrality node initializes the fog formation by creating the coalition subset $s \subset S$ as in Line 4. It adds its associated utility and cost (Lines 5 and 9) to the corresponding coalition utility and cost, respectively. For each content $x$ cached at the node $v'$, it updates the content availability in the corresponding coalition $s$ as stated in Lines $6-8$. The first condition in Line 10 ensures the cost of nodes $f_s^C$ merged into coalition $s$ do not exceeds a threshold cost $f_s^{C_{th}}$ dedicated by the service provider.

Moreover, the content availability $A_v^x$ for all content should also meet the minimum requirements by the service provider for contents availability at different locations and time-slots. The high centrality node requests the other subscribed nodes in the location $l$ and time-slot $\bar{t}$ to join the coalition. Any other node $u' \in \mathbb{V}$, joins the the coalition $s$ if it gets a higher utility than it can get by itself, at the same time, not surpassing buffer limits. It subsequently merges into the coalition and adds its utility to the coalition as in Lines $12-14$. For each content $x$ cached at the node $u'$, the content availability (Lines $15-17$) and its associated cost (Line 18) is updated at the set $s$. Finally, the set of nodes in the coalition $s$, their respective cached content availability $A_s^x$, their utility $U_s$, and cost function $f_s^C$ is returned (Line 23) to the service provider by the high centrality node in the location $l$ and time-slot $\bar{t}$.

The above distributed algorithm allows the nodes to forms a distributed fog in each location and time-slot, where, co-located nodes both spatial (in a location $l$) and temporal (within time window $\bar{t}$) join coalition as a stable set $s$, thereby not affected by mobility while in the same location and time window. Theorem 1 states that the existence such fogs can be combined to form a city-wide grand coalition as the set of coalitions $S \in \mathbb{V}$. The advantage of the grand coalition is that both the user and the service provider can benefit from the content cached at the large size virtual cache as a stable set of all the subscribed node with storage, thus maximizing the content availability near end users.

*Complexity and Convergence:* Algorithm 1 requires the nodes to only compute their respective utility which is a linear relation of its reward and cost function. It then shares this local information with the service provider and neighboring nodes where to avoid network overhead, a service provider can request a high utility node in a neighborhood to initiate the fog (coalitions) formation. The nodes are not required to know the global network knowledge, thus, resulting in substantially a reduced complexity algorithm.

The algorithm converges when the utilities of all the coalitions at each location and time-slot are maximized. This is possible when (i) the content availability requirements in the mobile fogs for each location and a time-slot are fulfilled, and, (ii) all the nodes subscribed for caching content form coalitions in the respective location and time-slot also referred as the "grand coalition" in the network.

## V. Performance Evaluation

### A. Use Case: Vehicles as Mobile Fogs

We evaluate the proposed coalition game model using ns-3 where the named-data networking model of the CCN architecture is implemented with a discrete time granularity of 1 second. The node mobility is extracted from a realistic model for large scale vehicular mobility in Cologne, Germany [15]. We choose vehicles as potential candidates for to form urban fogs as the relatively high computing, caching and communication capabilities of modern vehicles makes best candidates to form city-wide mobile fogs [16]. For the impact of fading and shadowing due to obstacles such as buildings, we use the combination of Nakagami and Log distance path loss model. A total of $2,986$ nodes are used to validate the scalability of our caching approach at mobile fogs. The analysis is performed by dividing an area of $6X6km^2$ into 36 uniform neighborhoods (fogs) each of $1X1km^2$, where, $l = 36$ and $\bar{t} = 10$ minutes.

### B. Simulation Scenario

The simulation scenario implements two types of nodes, the content caching or source nodes and the consumer nodes interested in the content. Consumer nodes generate interests for a pre-known content sequence of 100 chunks of size $1kb$ each following a Zipf distribution to model the content popularity profile allowing frequent interests for more popular content. Any caching node already containing the content responds to the consumer interest. We allow intermediate nodes to perform in-network caching where uniform buffer sizes are assigned to each node. $30\%$ consumers, $30\%$ caching nodes and the remaining nodes are intermediate nodes with no caching to better evaluate caching nodes. The content availability is modeled by randomly assigning it to each node as its cached content. The content validity metric $\tau = [0, 1]$ is extracted from its popularity as well as its availability profile by tracing the generated interests to compute Equation 1.

For each node, the centrality is computed using Equation 3 and the corresponding node utility is derived from its centrality, reward and cost extracted from its buffer space. Utilities for spatio-temporally co-located nodes are computed towards the respective coalition formation. Algorithm 1 is implemented to find the best coalitions as mobile fogs from the set of all coalitions formed in the network. We use STRIVE [17] as a social aware protocol to forward the user interests to retrieve content from caching nodes.

We perform each simulation up to ten times by analyzing different set of nodes as content caches in order to compute up to $95\%$ confidence intervals. The Fogs formed using the proposed approach is compared with the following schemes:

- Centrality-based Fog: Benchmark social-aware approaches use Degree, Closeness, Betweenness, Eigenvector centrality, thus we implement each centrality scheme to form fogs using Algorithm 1.
- No Fog: approach where no coalition is formed and each node greedily implement individual caching policy LRU in a non-collaborative fashion.

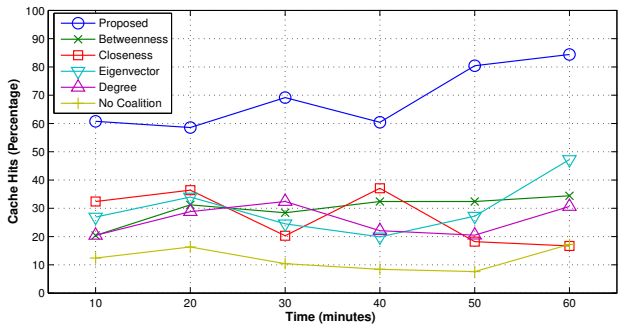

Figure 2: Average cumulative cache hit rate by all nodes with utilities computed using each scheme in ten simulations

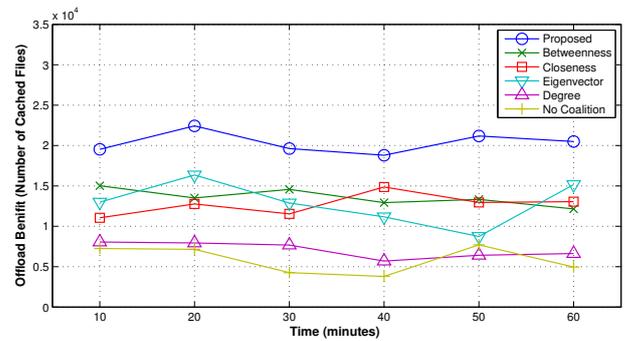

Figure 3: Total number of files cached using each scheme over an average of ten different simulations

The following performance metrics are used for evaluations:
- Cache Hits: Ratio of content responds from node cache for the received consumer interests.
- Offload Benefit: The amount of content files downloaded/retrieved from local nodes to relieve infrastructure, while nodes utility proportional to the content delivery is taken into account.
- Spatio-temporal Coalitions: Graphical depiction of the mobile fogs formed in the network in order to validate the coalition formation.

### C. Simulation Results

*1) Cache Hits:* We evaluate the overall cache performance by finding the total number of cache hits by the caching nodes in the network. Figure 2 depicts the cache hit rate in percentage by comparing different schemes obtained at 10 each minutes. This analysis clearly shows that the coalitions formed at the nodes using our scheme outperforms existing schemes with a minimum of $60\%$ and up to $85\%$ cache hits. Moreover, the case of no coalition between nodes resulted in the poorest performance in terms of cache hits reaching less than $10\%$ during the entire simulation. Existing social metrics such as typical centrality schemes yielded a cache hit rate of around $20-30\%$, thus, validating the efficiency of using our proposed scheme.

We also observe an increase in the cache hits over time for our approach. It is because the coalition formed at the nodes with high utility caches more content over time and subsequently satisfies more interests for the incoming requests. Such increase is not observed by the other schemes since each time slot at different location yields different set of nodes, thus cannot be relied for content caching. Most importantly, this also validates the fact that coalitions improves caching performance as all the cases where coalitions were formed (i.e. degree, betweenness, closeness, eigenvector based coalition) resulted in higher cache hits than the case where no coalition is formed.

*2) Offload Benefit:* The benefit of offloading at mobile fogs is validated by finding the total number of files cached in the mobile fogs. Figure 3 shows the offload benefit at different times in the simulation with shown after each 10 minutes. It is clear that the total number of content files made available for the user by caching at nodes by our approach is higher than with all existing schemes as well as in the case of no coalition. The case of no coalition failed to cache files substantially in the network with five times lesser number of files, where other schemes cached half the amount of files compared to our approach. This proves that caching content at mobile fogs, formed around high centrality nodes identified by our approach maximizes the overall content availability in the network. It is because such nodes received, cached and thus satisfied more user relevant interests in the network. The formed coalitions not only reduced the network overhead with respect to the number of files but also considered the content popularity to relieve the infrastructure network.

*3) Spatio-temporal coalitions:* It is important to evaluate the existence and uniqueness of the core in the proposed coalition game. Therefore, Figure 4 shows the mobile fogs formed by nodes, vehicles in this case, around 10 minutes, 30 minutes and 60 minutes into the simulation for the Cologne City center ($6km^2$). The central node of each coalition is marked as '◦' and the remaining nodes in the coalition are marked as '∗' . First, we observe that all nodes form coalition resulting in the existence of grand coalition as the core of the game. Most of the nodes are isolated nodes and thus form fog in the form of self-coalition as shown by a high number of single superimposed '◦' and '∗'. Secondly, we noticed that most coalition central nodes are located at the edge of the network. This validates the existence of mobile fogs in closer proximity to consumers. We also observe that a single central node represents the coalition of the most denser part (relatively center) of the network. This proves the efficiency of our game motivating nodes to join coalition in order to form a grand coalition. The network analysis at three different snapshots highlighted the benefit of considering spatio-temporal factor in our approach as we can see different set of coalitions are formed at different times. Thus, we can clearly observe the union of smaller coalition in different neighborhoods forming the grand coalition in the network.

### D. Summary of Findings

Results shown that offloading content can be efficiently managed by forming mobile fogs allowing nodes to perform

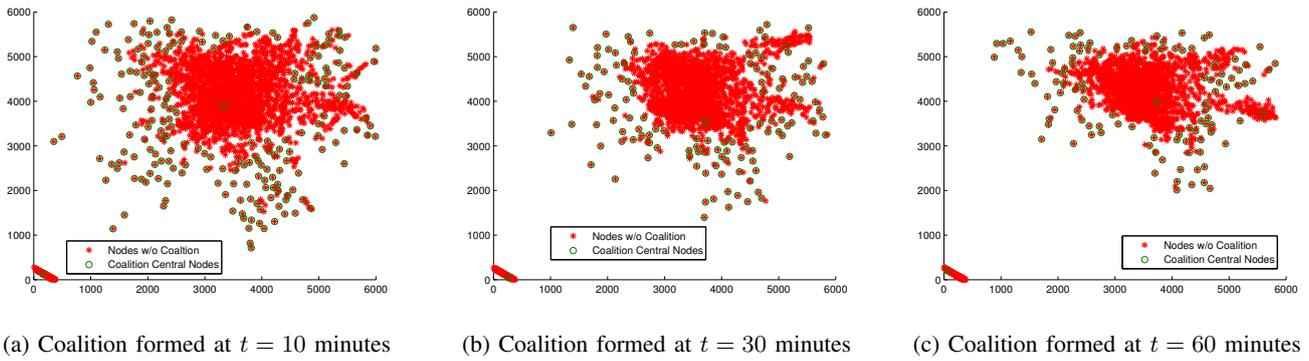

(a) Coalition formed at $t = 10$ minutes  (b) Coalition formed at $t = 30$ minutes  (c) Coalition formed at $t = 60$ minutes

Figure 4: Spatio-temporal Mobile Fogs for $6X6$ km area, central nodes for each coalition identified in green

distributed resource pooling. It is better to dedicate "high centrality" nodes for the distributed management of such coalitions in order to reduce overhead at the network infrastructure. As indicated in Figure 4, it is important for individual nodes to consider the spatio temporal content profile while considering both its availability and popularity towards caching decisions. We observe that in case of absence of other nodes in the vicinity, the node can also form self-coalition or self-fog as the only node caching content in the location. We also found that most coalition central nodes reside at the edge of the network which simultaneously reduce back-haul network traffic and improve user quality of experience. Thus, our analysis based on results from scalable simulation scenarios clearly validates our proposition that offloading content at mobile fogs in the form of coalitions at the network edge substantially benefits the network.

## VI. Conclusions and Future Directions

This paper proposed a collaborative content caching approach at high centrality nodes in an urban environment. To do so, we first modeled the content profile by defining its spatio-temporal popularity and availability relation allowing each node to find the content eligibility, then it finds its capability to cache content taking into consideration the offered resources and cost with respect to social interest of users. We proposed a coalition game for the best co-located candidates to self-organize and share storage close to mobile users. We performed scalable simulations for an interesting case of vehicles as mobile caches on a realistic trace. Results shown that the our proposed coalitions offload twice more content and achieved $60\%$ cache hits compared to other approaches which yielded around $25\%$ cache hits in the network.

Caching at mobile nodes need further investigations. Once we identified mobile fogs, the study of how to route user interests to nearby high centrality nodes for its efficient retrieval in the fog would be of interest. Therefore, our future work includes the study of content and node profile towards efficient content distribution in delay sensitive dynamic networks with multiple content providers while avoiding duplicate content downloads in the network.